\documentclass[10pt]{article}
\pdfoutput=1
\usepackage[usenames]{color} 
\usepackage{amssymb} 

\usepackage{amsmath} 
\usepackage{float}
\usepackage[pdftex]{graphicx}
\usepackage[utf8]{inputenc} 
\usepackage{geometry}
\usepackage[numbers,sort]{natbib}
\usepackage{url}
\usepackage{hyperref}

\begin{document}

\pdfoutput=1

\title{Beam Imaging and Luminosity Calibration}
\author{}
\maketitle
\begin{center}

Markus Klute, Catherine Medlock, Jakob Salfeld-Nebgen

Massachusetts Institute of Technology

\end{center}
We discuss a method to reconstruct two-dimensional proton bunch densities using vertex distributions from LHC beam-beam scans. The $x$-$y$ correlations in the beam shapes are studied and a luminosity calibration technique is introduced. The method is demonstrated using simulated beam-beam scans. We evaluate the impact of non-factorization features for beam shapes of double-Gaussian type.
 
\section{Introduction}
Absolute luminosity measurements at a particle collider are mandatory to determine absolute cross-sections for reaction processes. Specifically, the number of interactions, $N$, measured in an experiment depends on the value of cross-section $\sigma$ and of the luminosity $L$, with $N = \sigma L$. The uncertainty in the measurement of a given cross-section depends critically on, and is in some cases dominated by, the precision with which the luminosity is known. The LHC experiments calibrate the scale of the luminosity with  van der Meer (vdM) scans~\cite{vanderMeer:296752,Rubbia:1025746,White:2010zzc,CMS:2013gfa,Aad:2013ucp,Aaij:2014ida,Abelev:2014epa}. This technique, which involves scanning the beams across each other and monitoring the interaction rate, has been used by all of the four large LHC experiments. The method is intended to measure the overlap integral $O_{I}$ of the colliding proton beams with proton densities $\rho_1$ and $\rho_2$ 
\begin{equation}
\label{Aeff}
O_{I} = \int_{-\infty}^{\infty}\rho_1(x,y)\rho_2(x,y) \,dx\,dy ,
\end{equation}
after integration over the longitudinal coordinate and time. 

If $N_1$ and $N_2$ are the number of protons in the two colliding bunches and $\nu_{\mathrm{rev}}$ the LHC revolution frequency, the absolute instantaneous luminosity can be computed from directly measured accelerator parameters:
\begin{equation}
\mathcal{L} = N_1 N_2 \nu_{\mathrm{rev}} O_I.
\end{equation}
The collisions rate $R$ measured by a luminometer is related to the luminosity $\mathcal{L}$ and the visible cross section $\sigma_{vis}$ by
\begin{equation}
R = \sigma_{vis} \cdot \mathcal{L}.
\end{equation}

The reconstruction of the transverse beam shape poses a challenging problem in the luminosity scale calibration procedure of the LHC experiments. The vdM scan method relies on the assumption that the bunch proton densities are factorizable in the scanning plane of the detector, i.e. that $\rho_i(x,y) = \rho_i(x)\rho_i(y)$.  In general, this assumption does not hold and introduces one of the leading systematic uncertainties for luminosity calibration measurements~\cite{CMS:2013gfa,Aad:2013ucp,Abelev:2014epa,Aaboud:2016hhf,alice:2160174}. 
The LHCb collaboration exploits the combination of beam-gas and beam-beam vertex distributions to reconstruct the individual proton bunch densities~\cite{Aaij:2014ida,Ferro-Luzzi:844569}. 
Another approach exploits the simultaneous evolution of the luminosity and of the position, size, shape and orientation of the luminous region~\cite{Aad:2013ucp,Aaboud:2016hhf,alice:2160174} during beam-beam scans. In addition, a dedicated tailoring of the proton bunches in the injection chain was developed to minimize non-Gaussian contributions to the transverse beam profiles~\cite{Bartosik:1590405}.

In this paper a method to estimate the $x$-$y$ correlations is developed and a new proposal for a complementary luminosity calibration is presented. The method generalizes the beam imaging technique proposed in~\cite{Balagura:2011yw} and experimentally realized in~\cite{Zanetti:1357856,Aaij:2011er} to two dimensions. 

In contrast to the standard vdM scan, one beam is scanned, first in $x$ and then in $y$, across the other beam, with the latter remaining stationary in the rest frame of the detector. 
The distributions of reconstructed collision vertices in the transverse plane accumulated during the scans constrain the two-dimensional proton-density distributions and are fitted simultaneously to determine the parameters of an analytical model for the proton densities of the two beams. As a result, $O_I$ can be computed and used to estimate the instantaneous luminosity including the $x$-$y$ correlations potentially present in the proton-density distributions.

This paper is organized as follows: In Section~\ref{vdmCorrStudy} the impact of correlations on the standard vdM scan method is studied. In Section~\ref{2dbeamimaging} a new approach to reconstruct the beam shapes is introduced and tested on simulated beam-beam scans. In Section~\ref{biasregression} a potential bias on the beam overlap estimation is studied and a correction based on a specific regression is applied. The conclusions are summarized in Section 5.

\section{Impact of X-Y Correlations on the vdM Standard Analysis}
\label{vdmCorrStudy}

In this section the impact of $x$-$y$ correlations in the LHC bunch proton densities on the standard vdM scan method is demonstrated. A set of simulated vdM scans is generated, based on the double-Gaussian single beam density distributions $b_{i}$ for beam $i$ ($i = 1, 2$) :
\begin{eqnarray}
\label{doubleGauss}
b_i (x,y) = w_i g_{i,N}(x,y) + (1-w_i) g_{i,W}(x,y).
\end{eqnarray}
Each beam is a sum of a "narrow" Gaussian component $g_{i,N}$ and a "wide" Gaussian component $g_{i,W}$, where each is of the following form:
\begin{eqnarray}
\label{modeldoubleGauss}
 g_{i,j}(x,y) = \frac{1}{2\pi \sigma_{i,j,x} \sigma_{i,j,y} \sqrt{1-\rho_{i,j}^2}} \exp{\bigg( \frac{-1}{2(1-\rho_{i,j}^2)}\Big[  \frac{x^2}{\sigma_{i,j,x}^2}+\frac{y^2}{\sigma_{i,j,y}^2}-\frac{2\rho_{i,j} x y}{\sigma_{i,j,x}\sigma_{i,j,y}}\Big]\bigg) }.
\end{eqnarray}
The $x$-$y$ correlations, or non-factorizability of the proton beam densities, are parametrized by the two weights $w_i$ and the four correlation parameters $\rho_{i,j}$ ($i = 1, 2$, $j = N, W$).

For each element in this set, the one-dimensional $x$- and $y$-scan curves are fitted with a double-Gaussian model and the beam overlap integral $O_{I}$ is computed based on the assumption on the factorizability in $x$-$y$ of the bunch proton densities
\begin{align}
\label{overlapVdMfact}
O_I = \int_{-\infty}^{\infty} \rho_1(x)\rho_2(x) \,dx \times \int_{-\infty}^{\infty} \rho_1(y)\rho_2(y) \,dy.
\end{align}

\begin{figure}[H]
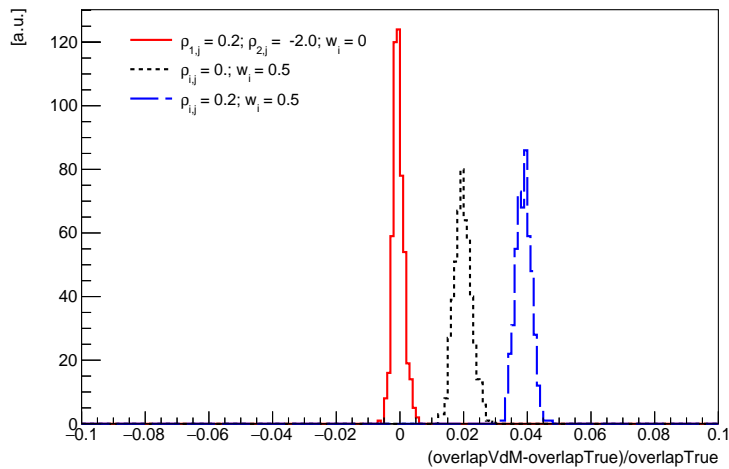

\centering
\includegraphics[width=300bp,height=195bp]{{{corrCompNEW2}.pdf}}
\caption{An estimate of the impact of $x$-$y$ correlations described in Equations~\ref{doubleGauss} and ~\ref{modeldoubleGauss} on the vdM scan method is shown. The red histogram shows the relative difference between the reconstructed and true beam overlap integral for beam-shape parameters $w_i=0$, $\rho_{1,j}=0.2$ and $\rho_{2,j}=-0.2$. The black histogram shows the difference for  beam-shape parameters  $w_i=0.5$ and $\rho_{i,j}=0$, and the blue histogram shows the relative difference for $w_i=0.5$ and $\rho_{i,j}=0.2$.\label{correlationsVdM}}
\end{figure}

To illustrate the impact of $x$-$y$ correlations in the beam shapes, the relative difference between the true overlap integral and the one extracted from simulated vdM scans is shown in Figure~\ref{correlationsVdM} for three different beam-shape samples.
For each beam-shape type the widths are varied within $\sigma_{N,x,y} \in [1.6, 1.8]$ and  $\sigma_{W,x,y} \in [2.4,2.6]$, in arbitrary units. The red histogram shows the case for which $w_i=0$, $\rho_{1,j}=0.2$ and $\rho_{2,j}=-0.2$, the case for which $w_i=0.5$ and $\rho_{1,j}=0$ is shown in the black histogram, and the blue histogram shows the impact of choosing $w_i=0.5$ and $\rho_{i,j}=0.2$. In these examples, the bias is largest for correlation parameters having the same sign and $w_i=0.5$, and reaches up to 4\%. A bias in the estimation of the overlap intergral leads directly to a bias in the luminosity measurements .

\section{Two-dimensional Beam Imaging}
\label{2dbeamimaging}
As was shown in the previous section, the vdM-scan method may result in a significant bias, if the bunch proton densities exhibit $x$-$y$ correlations. Therefore, sizeable systematic uncertainties were estimated and assigned by the CMS and ALICE Collaborations~\cite{CMS:2013gfa,Abelev:2014epa} and non-factorization effects of up to 4\% were found by the LHCb~\cite{Aaij:2014ida} and~\cite{Aaboud:2016hhf} ATLAS Collaborations.

A higher precision may be achieved by fully reconstructing both two-dimensional proton density distributions of the colliding bunches. A natural choice is to measure the vertex distributions in the transverse plane, the density of which is proportional to the product of the proton densities in the two colliding bunches. The density of the number of vertices at a given point $(x,y)$ accumulated during a time interval $\Delta T$ and convolved with the vertex-position resolution $V$ is given by
\begin{align}
\label{numVertices}
n^{vtx}(x,y) = \rho_1(x,y) \rho_2(x,y) \, \left( \nu_{rev} \Delta T N_1 N_2 \sigma_p \epsilon^{vtx}\right) \otimes V,
\end{align}
where $N_i$ ($\approx 10^{11}$) is the number of protons in the bunch considered, $\nu_{rev}$ is the revolution frequency, 11245 Hz, for a bunch in the LHC ring, $\epsilon^{vtx}$ is the vertex reconstruction efficiency and $\sigma_p$ is the proton-proton cross section. 
During vdM scans, the experiments record collision data at very high rate, using an unbiased trigger on selected colliding bunches. For the study in this paper, we assume about $10^6$ reconstructed vertices per scan point. During the Run-1 vdM scan campaigns~\cite{Aaij:2014ida}, beam optics were chosen such that the transverse luminous size yields $\sigma_\mathcal{L}\approx 60$ $\mu m$,  corresponding to normalized transverse emittance $\epsilon_N\approx 3 \cdot 10^{-6} m$ and $\beta^*\approx 11m$ and a single beam width of
\begin{equation}
\label{beamwidth}
\sigma_b = \sqrt{\epsilon_N \beta^*/\gamma} \approx 90 \mu m.
\end{equation}
For comparison, the vertex-position resolution $V$ in the CMS Run-1 detector varies from 20 to 100 $\mu m$, depending on the transverse momentum and the number of vertex-associated tracks~\cite{Chatrchyan:1704291}. In this paper, an average resolution of $\sigma_b\approx 30$ $\mu m$ is assumed, which with the beam parameters of Equation~\ref{beamwidth} amounts to $\sigma_V \approx \frac{1}{3}\sigma_{b}$. The impact of uncertainties related to the vertex position resolution are discussed at the end of this section.

As described in the introduction an asymmetric scan setup is used as proposed in~\cite{Zanetti:1357856} and ~\cite{Balagura:2011yw}. 
To derive the fit model for the vertex distributions 
and without loss of generality the constants in Equation~\ref{numVertices} are set to 1
$$\nu_{rev} \Delta T N_1 N_2 \sigma_p \epsilon^{vtx} := 1.  $$
Considering a horizontal ($x$) scan of Beam~2 across Beam~1, in scan steps of size $\Delta x$, the following equation holds
\begin{align}
\sum_{n} n^{vtx}(x, y; n\Delta x) \Delta x & = \sum_{n} \rho_1(x,y) \rho_2(x+n\Delta x,y) \Delta x \otimes V\nonumber \\
 & =  \bigg[  \sum_{n} \rho_1(x,y) \rho_2(x+n\Delta x,y) \Delta x \bigg] \otimes V \nonumber\\
 & \approx  \bigg[  \int_{\Delta x} \rho_1(x,y) \rho_2(x+\Delta x,y) \,d(\Delta x)\bigg] \otimes V \nonumber\\
 & =  \rho_1(x,y) (\mathcal{M}_x \rho_2)(y) \otimes V .\label{deriveFitModelX1}
\end{align}
In the first step the distributivity property of convolutions is used and the approximation in the second step is the replacement of the sum over discrete scan points with a continuous integral over the beam-beam separation. After the integration, the $x$ coordinate is integrated out and the proton-bunch density of the moving beam appears marginalized in $x$.

Considering four scans of this kind, first scanning Beam~1 over Beam~2 in $x$ and $y$ and then vice versa, four two-dimensional vertex distributions are accumulated:
\begin{align}
n^{vtx}_{x,1} (x,y)= \rho_1(x,y) (\mathcal{M}_x\rho_2)(y) \otimes V \label{deriveFitModel3}\\
n^{vtx}_{y,1} (x,y)= \rho_1(x,y) (\mathcal{M}_y\rho_2)(x) \otimes V.\label{deriveFitModel4}\\
n^{vtx}_{x,2} (x,y)= \rho_2(x,y) (\mathcal{M}_x\rho_1)(y) \otimes V \label{deriveFitModel1}\\
n^{vtx}_{y,2} (x,y)= \rho_2(x,y) (\mathcal{M}_y\rho_1)(x) \otimes V \label{deriveFitModel2}
\end{align}

Each distribution constrains different parts of the underlying proton-bunch density distributions. For example, the $x$-dependence of $n^{vtx}_{x,2} (x,y)$ is determined by Beam~2 only (up to a scale factor), and therefore this distribution primarily constrains the horizontal shape of Beam~2 at different $y$-values. Furthermore, it contains $x$-$y$ correlations if and only if there are $x$-$y$ correlations in the proton-bunch density of Beam~2.
Assuming analytic models for $\rho_1(x,y)$ and $\rho_2(x,y)$, as well as the spatial vertex-position resolution $V$, the four distributions can be fitted and the full two-dimensional proton densities of the two colliding bunches of Beam~1 and Beam~2 are estimated.
For arbitrary beam-shape models the convolution with the vertex-position resolution is only possible by utilizing numerical convolution methods which are computationally intensive.
The application of deconvolution and unfolding methods in combination with the beam imaging in beyond the scope of this paper.

As an example, we apply the method to simulated beams of double-Gaussian shape, which in some of the previous vdM scan campaigns were observed to sufficiently describe the bunch proton densities~\cite{Aaij:2014ida}. The vertex reconstruction can be approximated by a sum of Gaussian resolution models. 
The fit model from Equations~\ref{deriveFitModel3} to~\ref{deriveFitModel2} is used and fitted simultaneously to the four two-dimensional vertex distributions. We use the ROOFIT~\cite{Verkerke:2003ir} package to perform the fits. Table~\ref{fitexampletable} shows beam imaging fit results for an example where the true beam overlap, the one reconstructed by beam-beam imaging, and that extraced from a conventional (i.e. factorizable) vdM analysis are, respectively:
\begin{align*}
O_I^{\mathrm{true}}  &= 0.0202, \\ 
O_I^{\mathrm{BI}} & = 0.0202 \pm 0.0001 ,   \\  
O_I^{\mathrm{vdM}}  &= 0.0206 \pm 0.0001 ,
\end{align*}
where the uncertainties are statistical-only.

\begin{figure}[H]

\centering

\includegraphics[width=180bp,height=173bp]{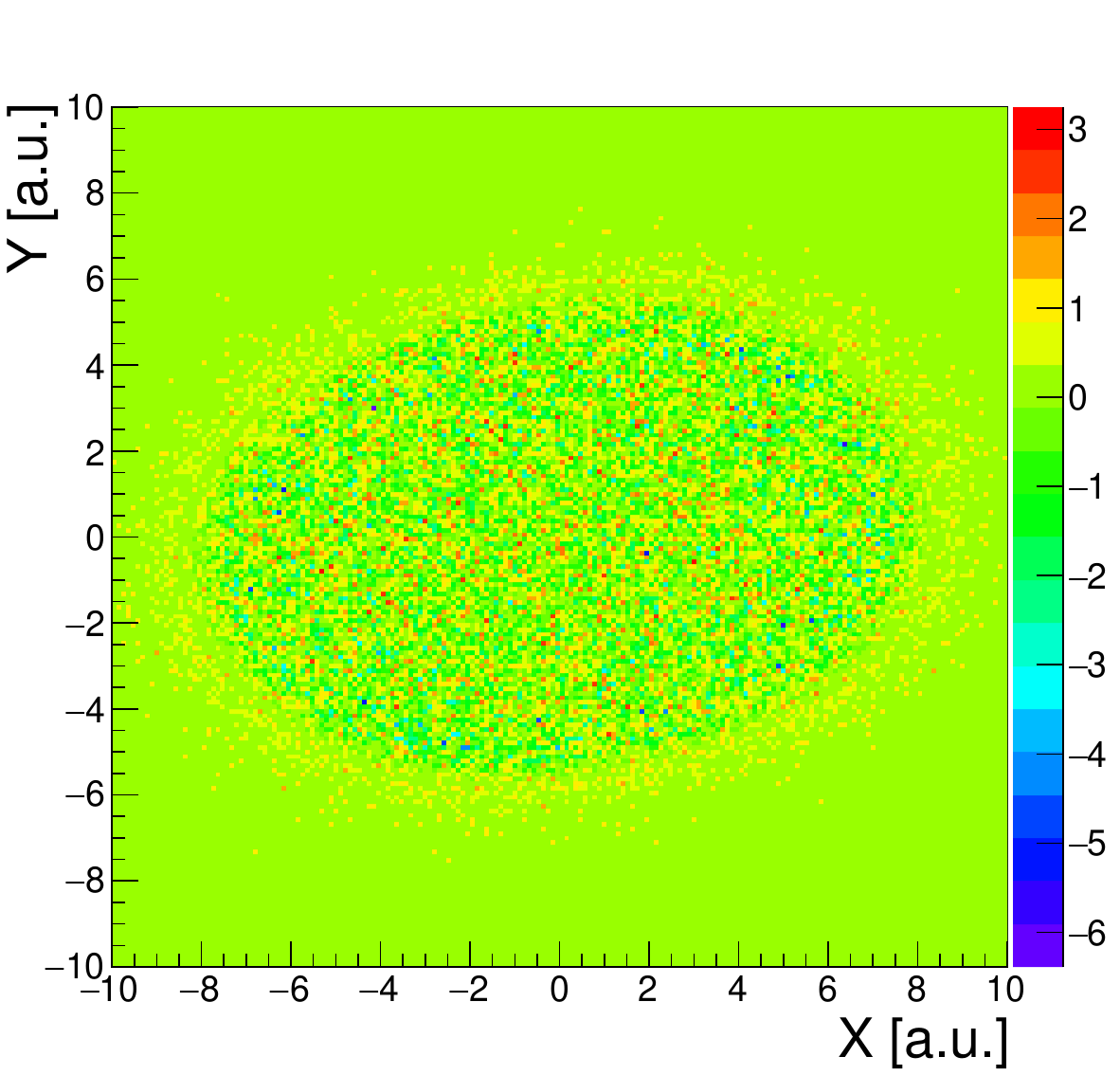}
\includegraphics[width=180bp,height=173bp]{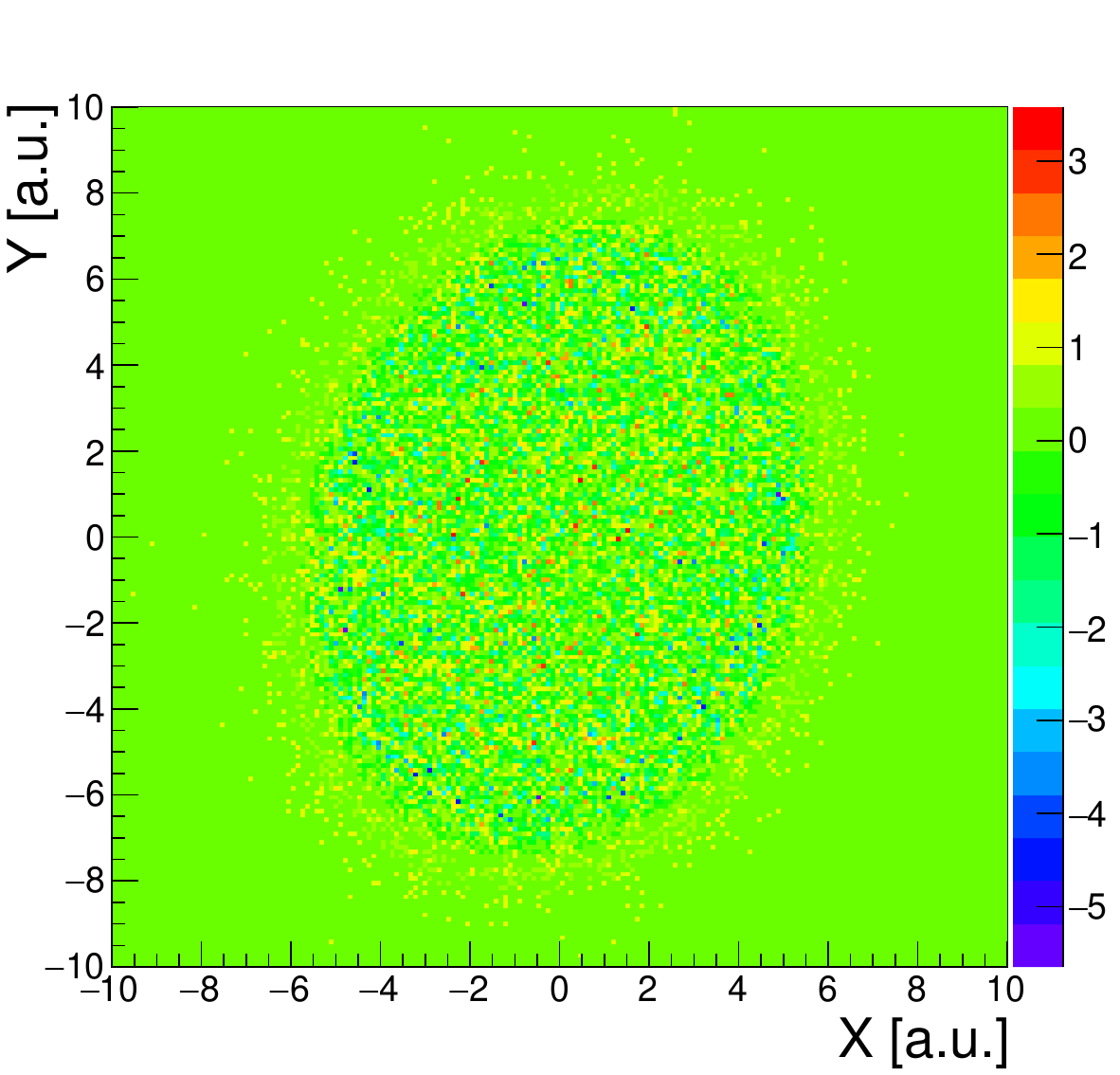}
\includegraphics[width=180bp,height=173bp]{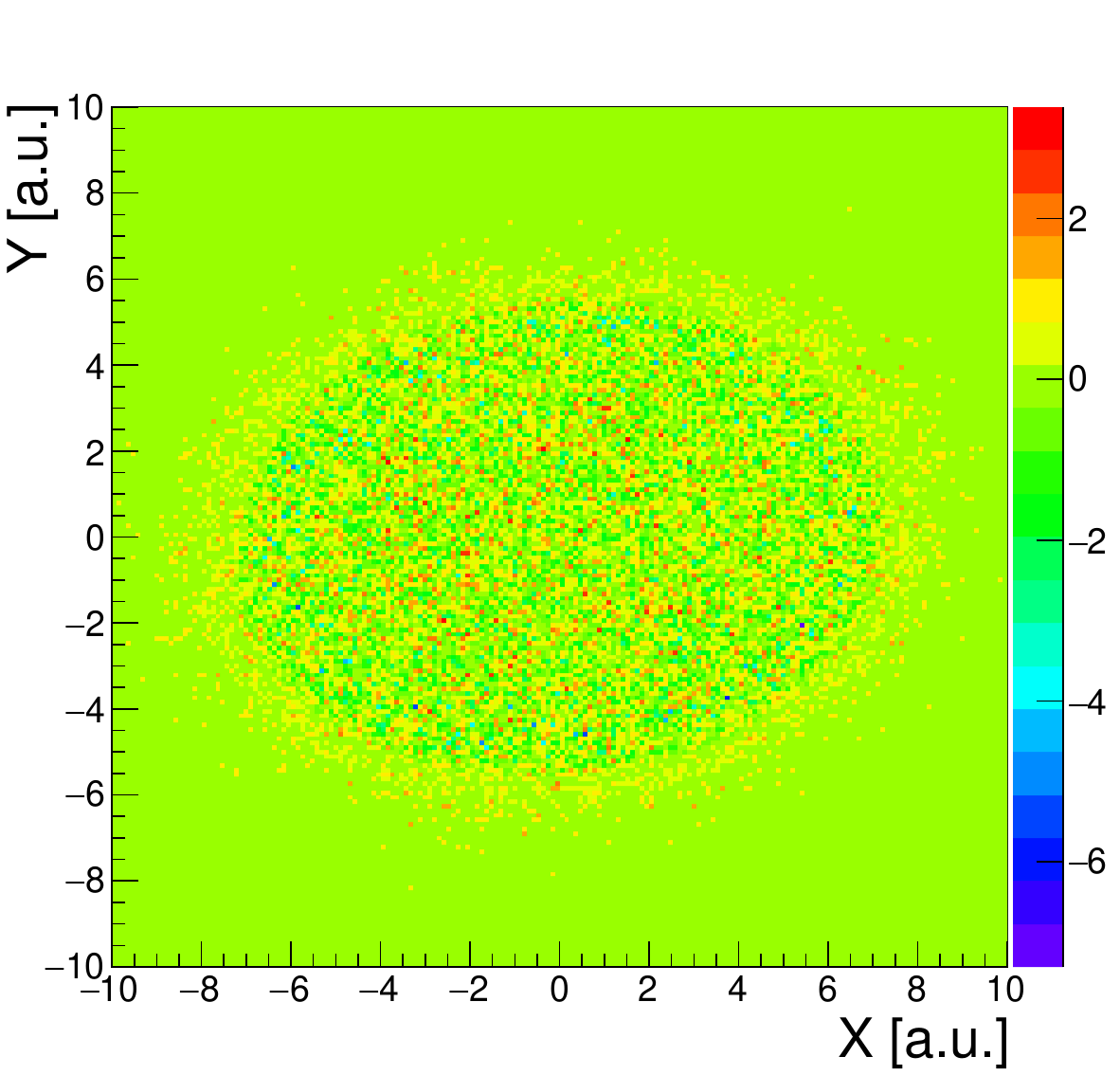}
\includegraphics[width=180bp,height=173bp]{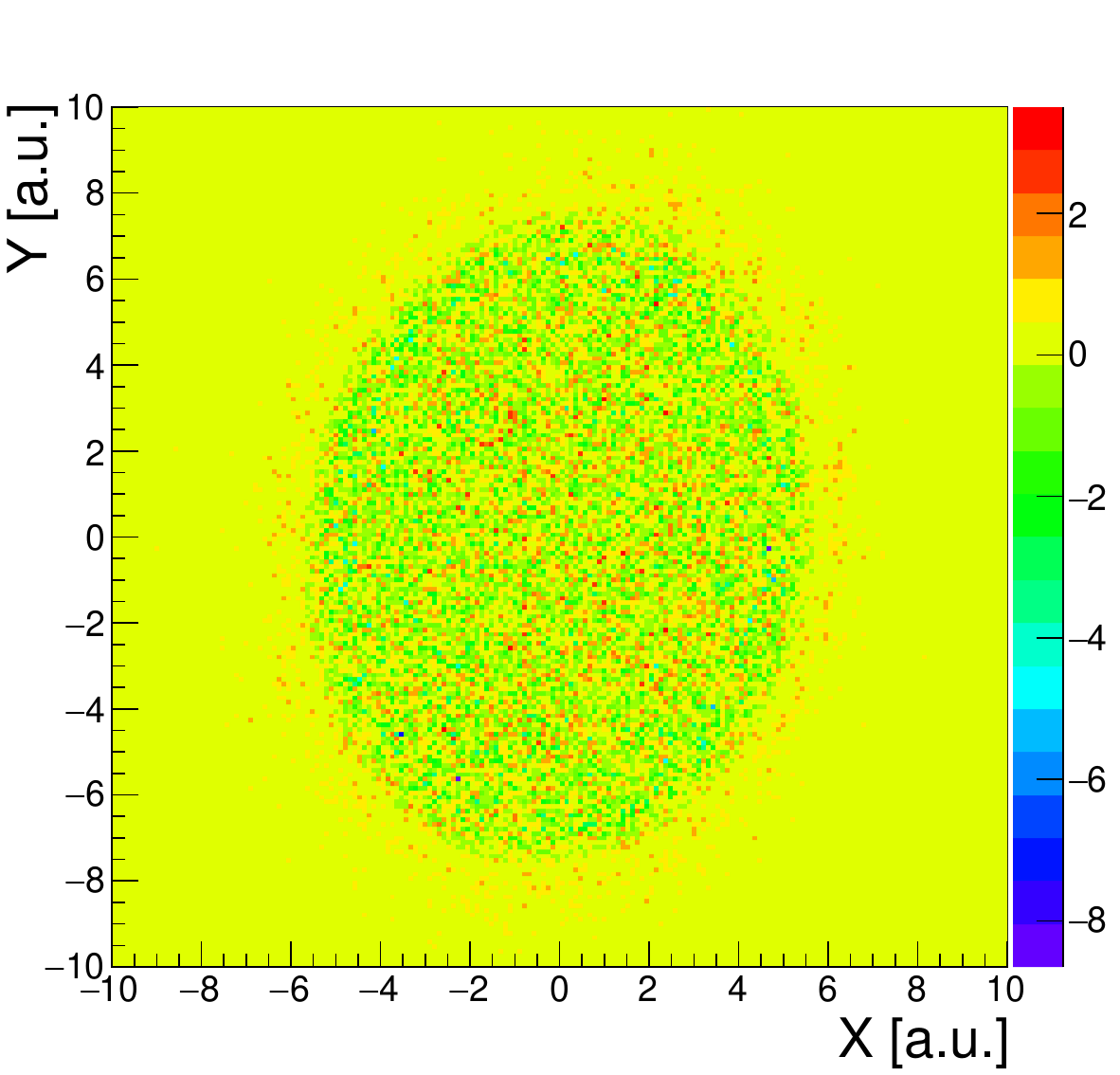}
\caption{Pulls of the two fitted double-Gaussian beam shapes to the four two-dimensional vertex distributions accumulated during simulated beam imaging scans. Beam~1 at rest and scanned with Beam~2 in $x$ (top left), Beam~1 at rest and scanned with Beam~2 in $y$ (top right), Beam~2 at rest and scanned with Beam~1 in $x$ (bottom left), Beam~2 at rest and scanned with Beam~1 in $y$ (bottom right). The best-fit parameters a shown in Table~\ref{fitexampletable}.\label{DGFit}}
\end{figure}

\begin{figure}[H]

\centering

\includegraphics[width=180bp,height=173bp]{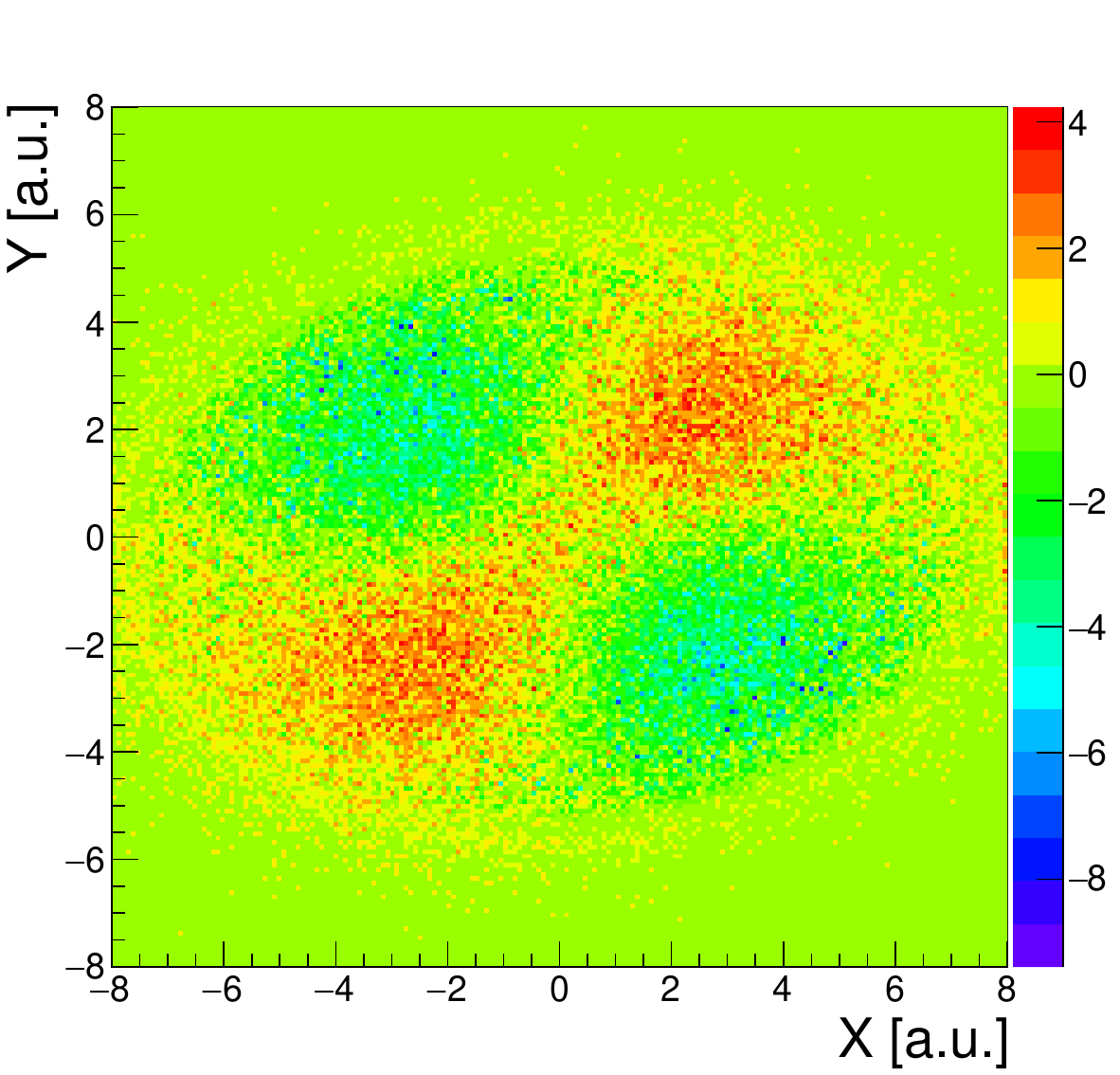}
\includegraphics[width=180bp,height=173bp]{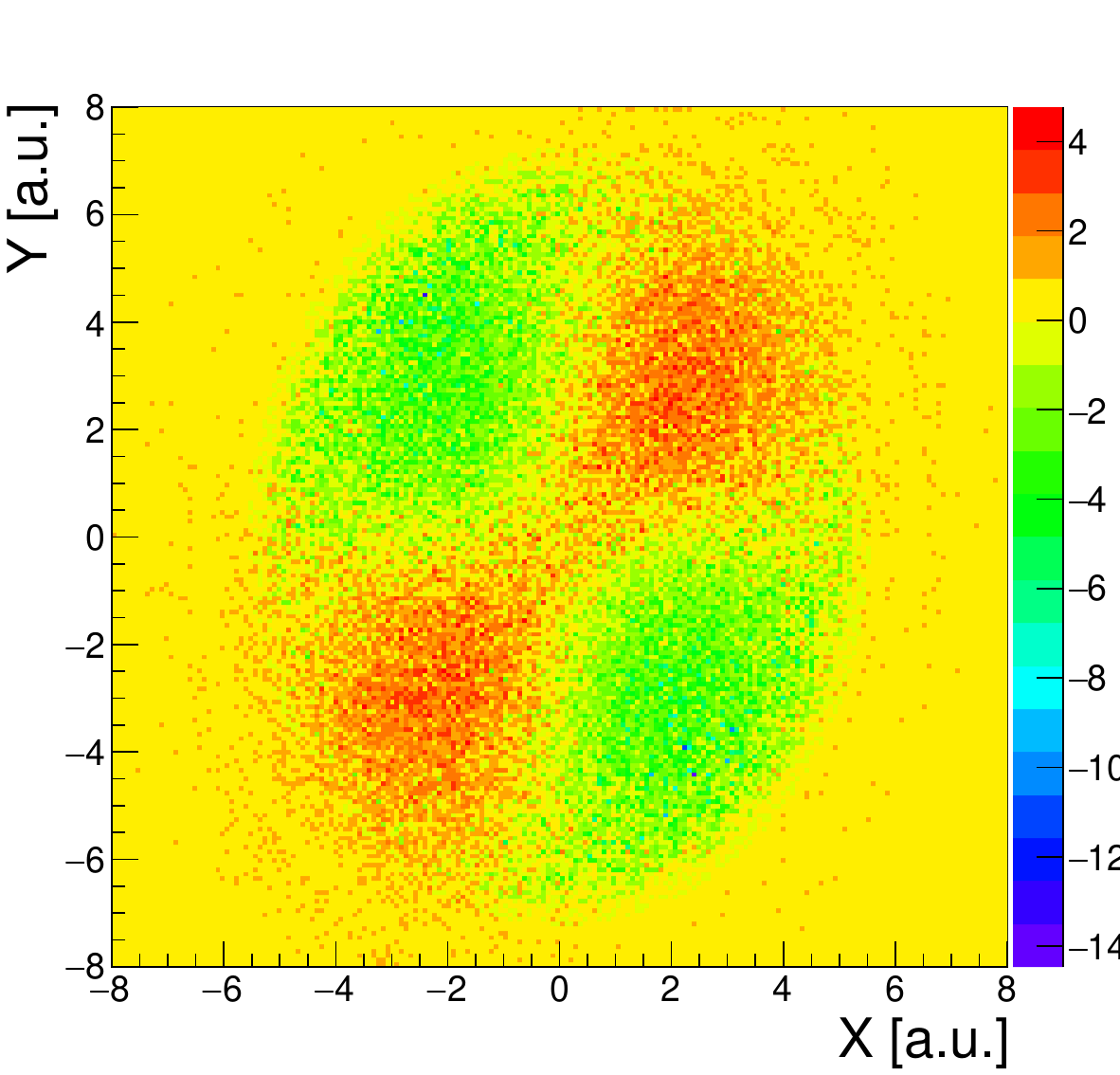}
\includegraphics[width=180bp,height=173bp]{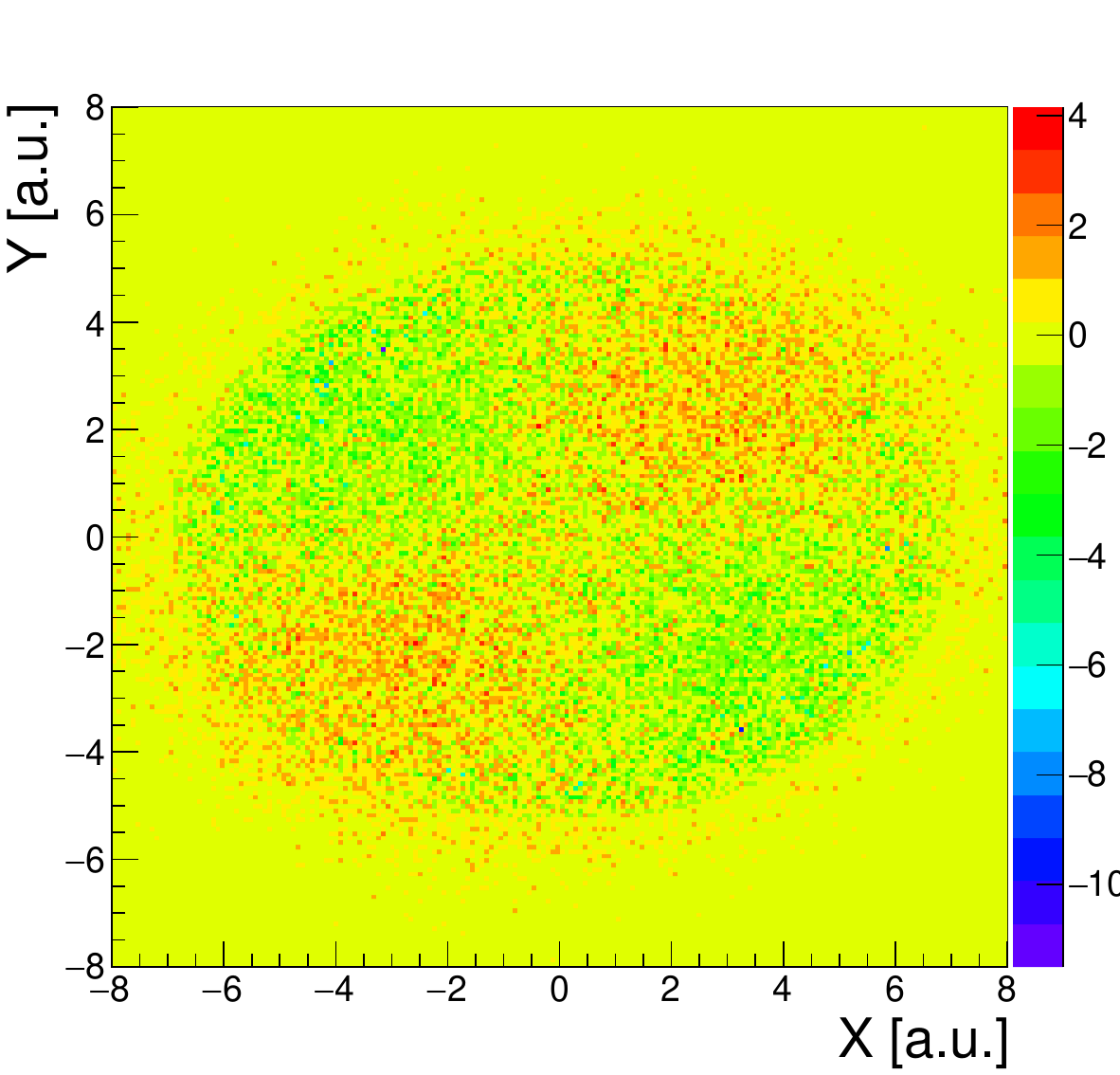}
\includegraphics[width=180bp,height=173bp]{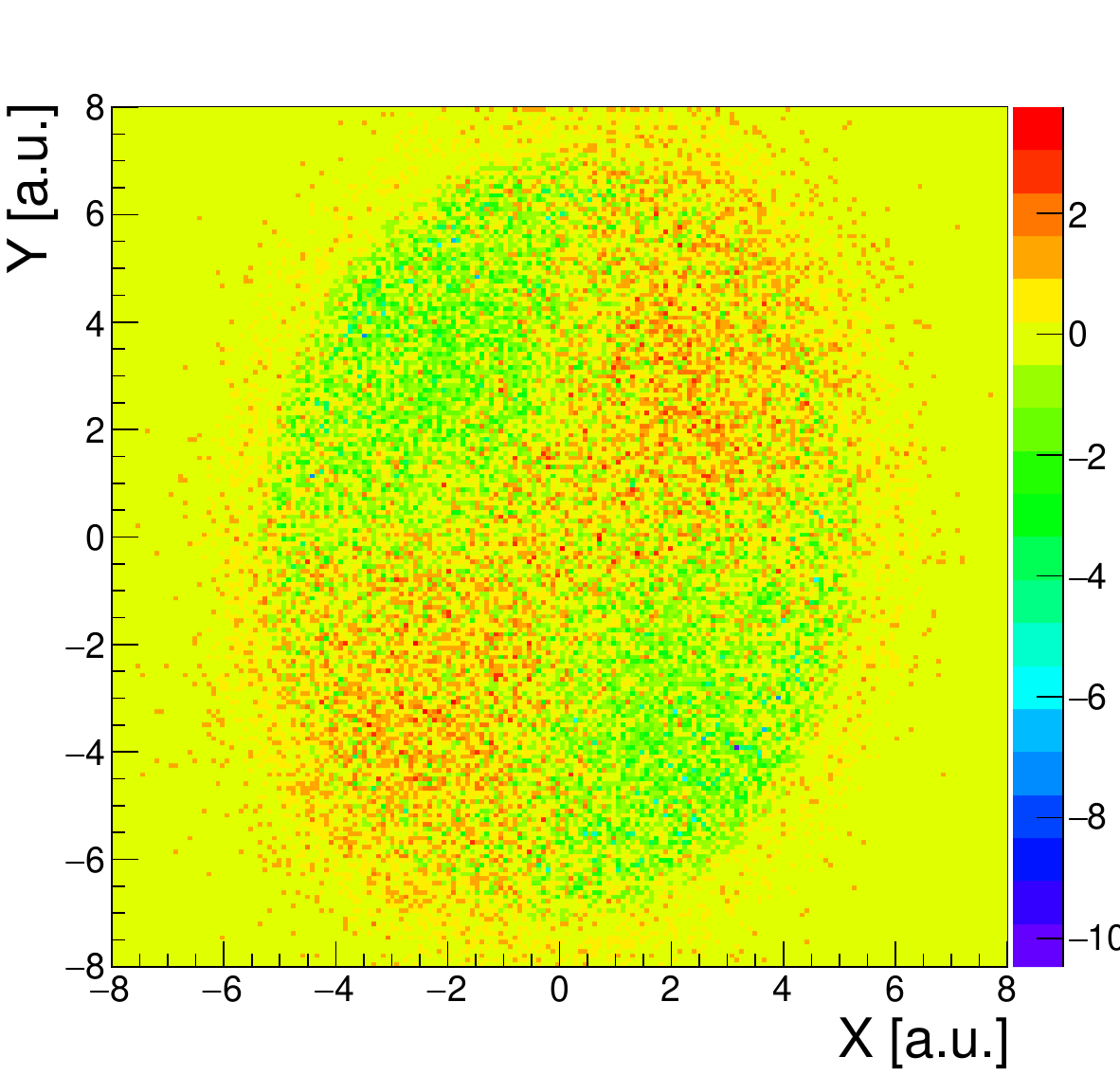}
\caption{Pulls of the two fitted single-Gaussian beam shapes to the four two-dimensional vertex distributions accumulated during simulated beam imaging scans. Beam~1 at rest and scanned with Beam~2 in $x$ (top left), Beam~1 at rest and scanned with Beam~2 in $y$ (top right), Beam~2 at rest and scanned with Beam~1 in $x$ (bottom left), Beam~2 at rest and scanned with Beam~1 in $y$ (bottom right).\label{SGFit}}
\end{figure}

While the beam imaging method measures a beam overlap consistent with the true beam overlap, a relative difference of about 2\% is observed when the vdM scan method is applied to measure the beam overlap. The pull distributions of the fitted double-Gaussian beam shapes to the four two-dimensional vertex distributions are shown in Figure~\ref{DGFit}. The $x$-$y$ correlations are visualized in Figure~\ref{SGFit}, showing the pull distributions of two single-Gaussian beam shapes fitted to the four two-dimensional vertex distributions. The reduced $\chi^2$ values for the fits using a double-Gaussian and single-Gaussian model as shown in Fig.~\ref{DGFit} and~\ref{SGFit} are $0.99$ and $2.11$, respectively.

\begin{table}
\centering
\begin{tabular}{c || c | c | c | c}
Parameter & Input Beam~1 & Fit Beam~1 & Input Beam~2 & Fit Beam~2 \\
\hline
\hline
   $\sigma_{W,x}$  &  2.358  & $2.319 $   &  2.235  & $ 2.045 $   \\
   $\sigma_{W,y}$   &  2.014  & $1.983  $  &   2.377  & $  2.141 $  \\
   $\sigma_{N,x}$  &  1.874  & $1.881 $    &  1.911   &  $ 1.886 $   \\
   $\sigma_{N,y}$  &  1.955  &  $1.981 $ &  1.932   &   $1.892$ \\
   $\rho_N$ &  0.395   &  $0.419 $ &   0.063    &  $0.011 $\\
   $\rho_W $ &  0.120   & $ 0.123$   &   0.302  &  $0.212$ \\
   weight  &   0.521  &   $ 0.482$    &  0.874   &  $ 0.609$ \\
\end{tabular}
\caption{Example of fit results for one simulated beam imaging scan for Beam~1 and Beam~2. True double-Gaussian parameters and the parameters extracted from the beam imaging fit procedure are shown. \label{fitexampletable}} 
\end{table}

\begin{figure}
\centering
\includegraphics[width=180bp,height=165bp]{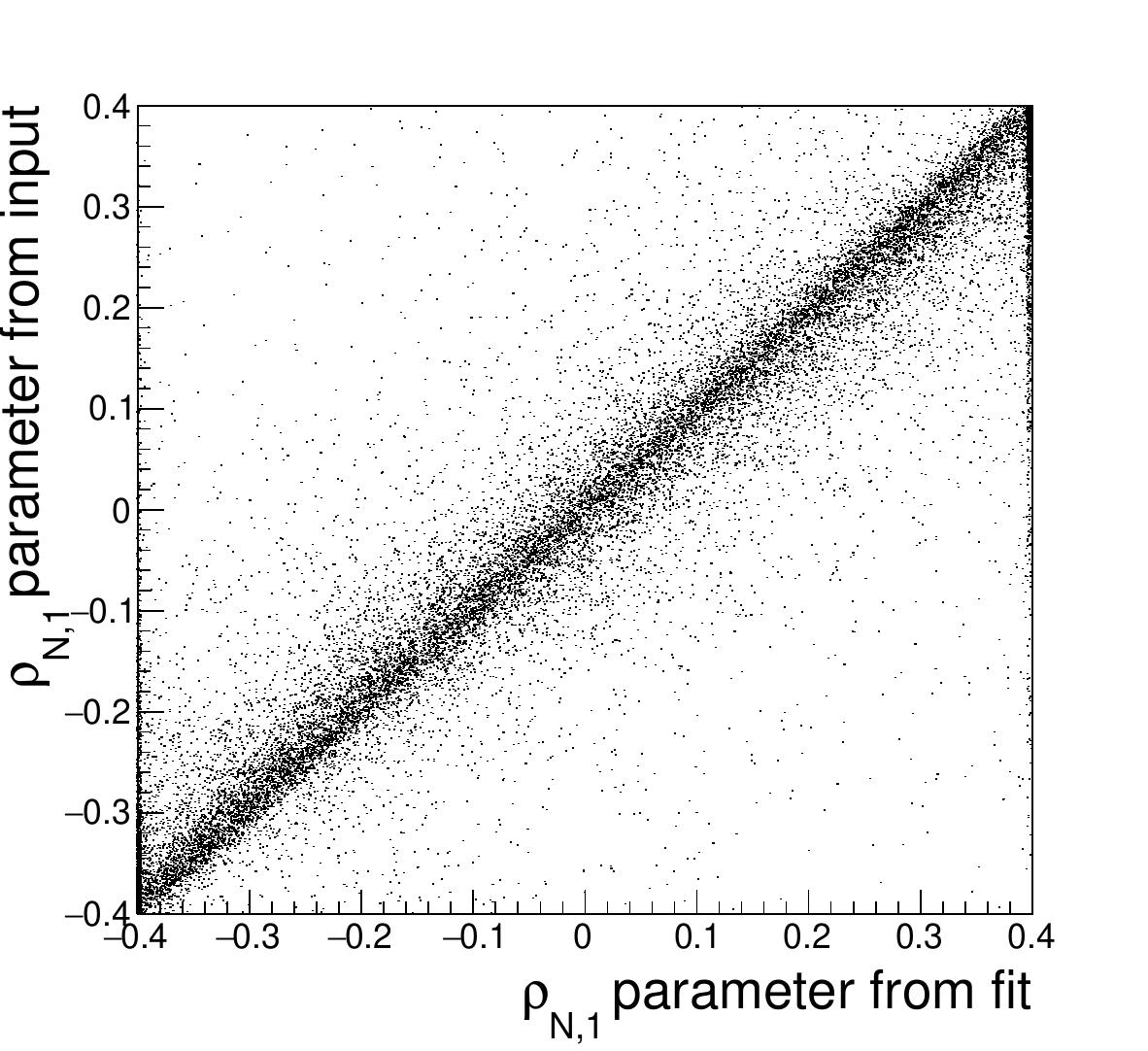}
\includegraphics[width=180bp,height=165bp]{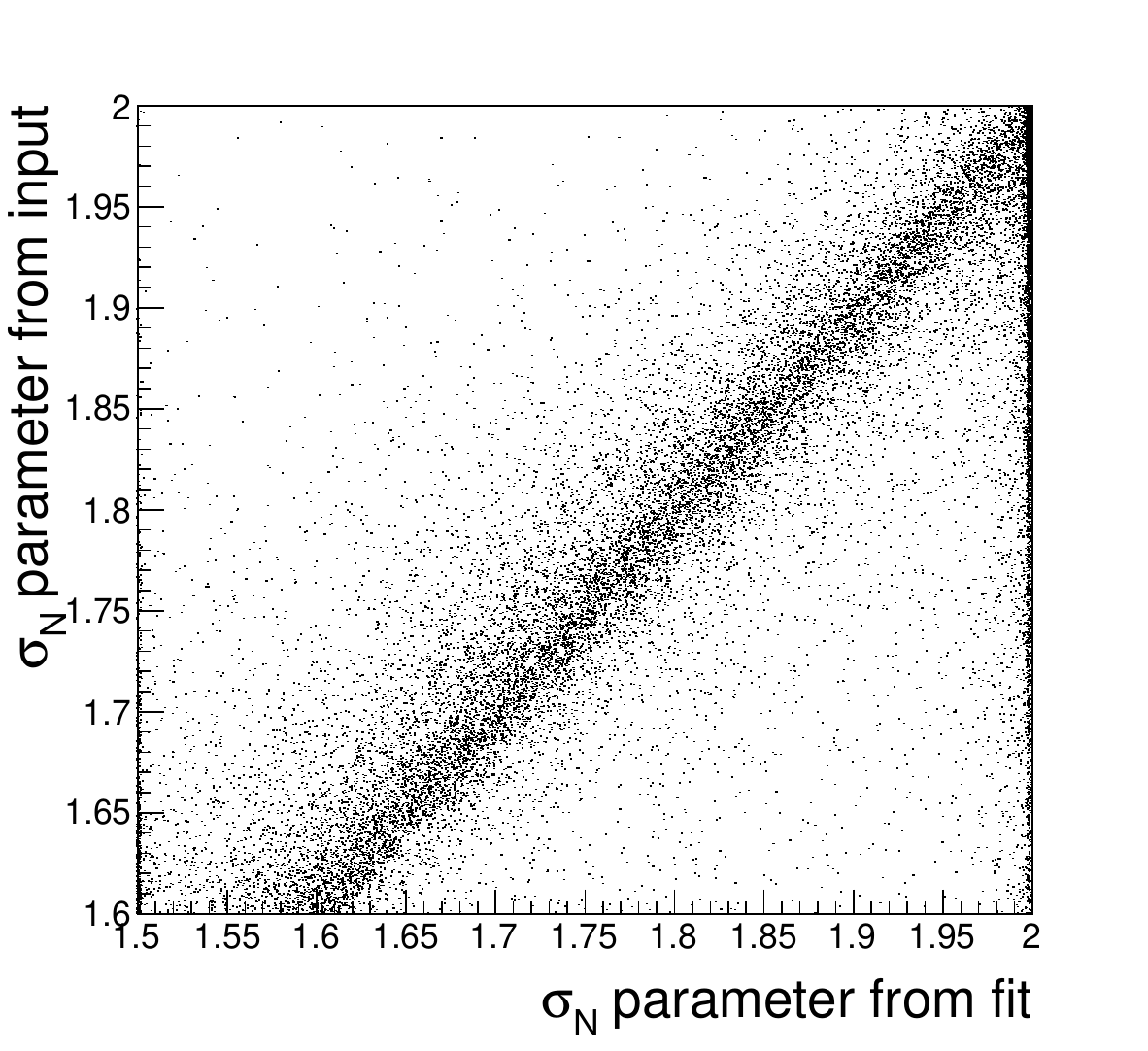}
\includegraphics[width=180bp,height=165bp]{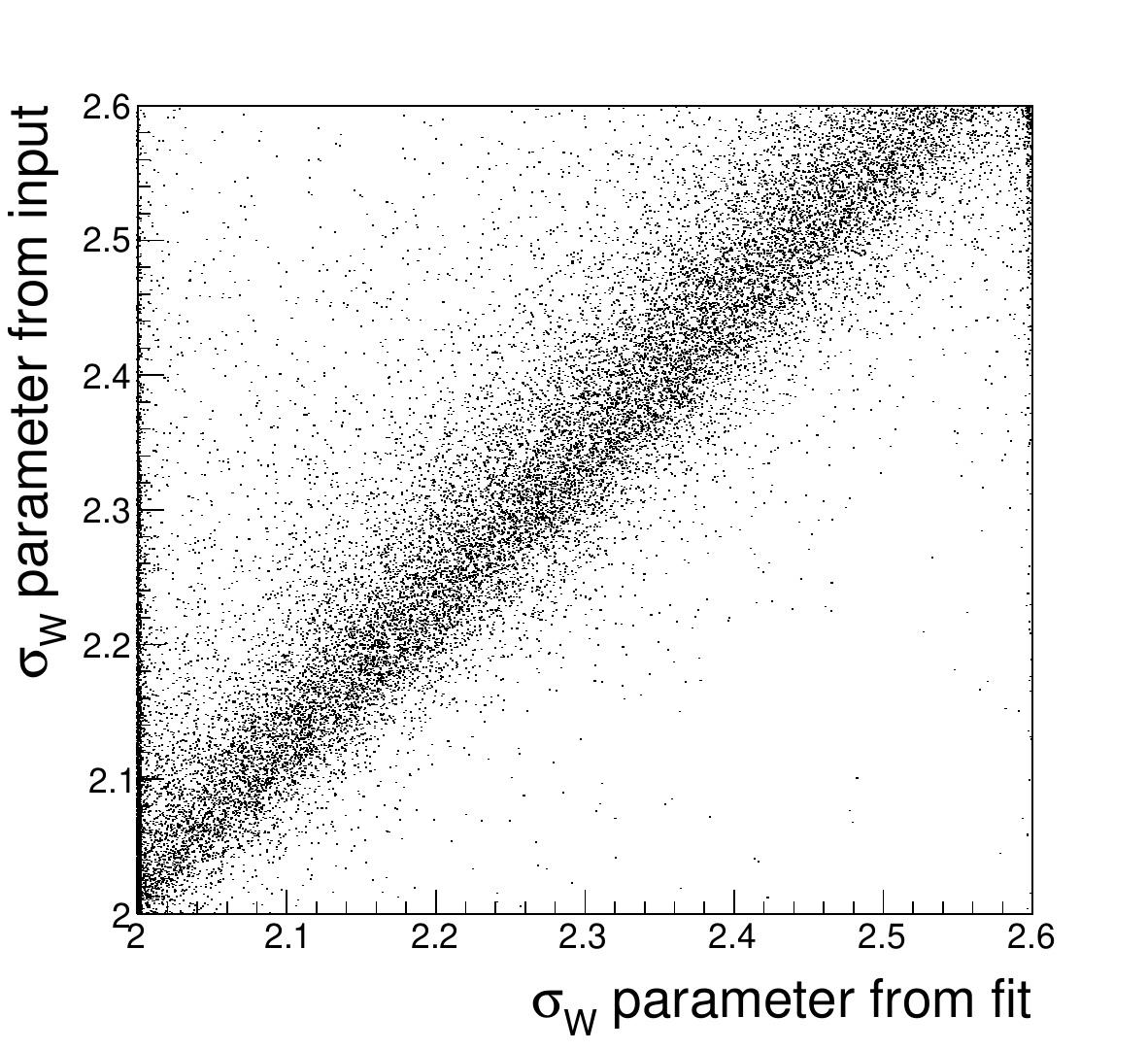}
\includegraphics[width=180bp,height=165bp]{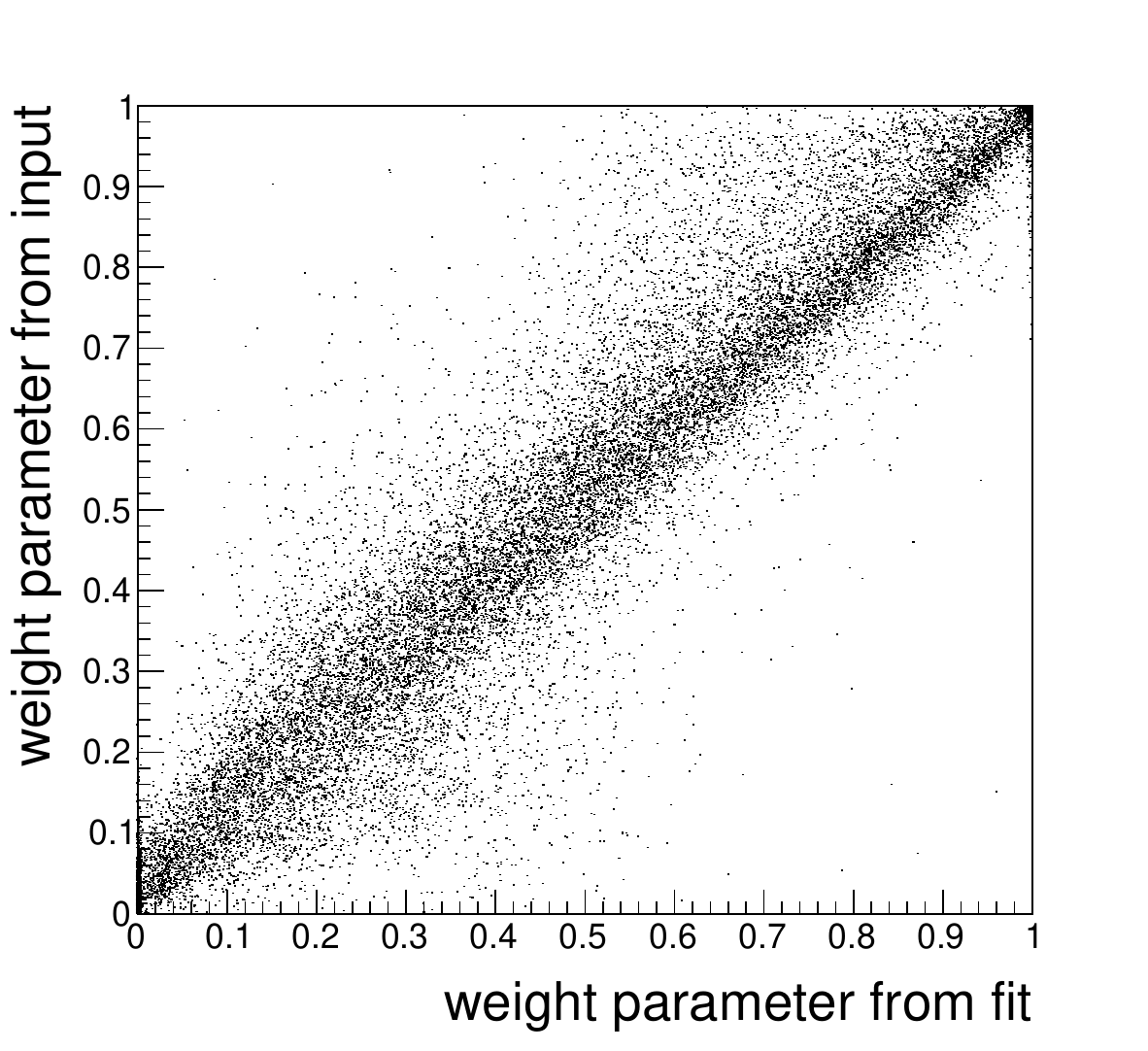}
\caption{Examples of correlations between the fitted ($x$-axis) and true ($y$-axis) beam shape parameters are shown. Top left shows the correlation between the correlation parameters for the narrow double-Gaussian component of Beam~1, top right shows the correlation between the width parameter for the narrow double-Gaussian component of Beam~1, bottom left shows the correlation between the width parameter for the wide double-Gaussian component of Beam~2 and the bottom right plot shows the correlation between the weight parameter for the double-Gaussian of fit model of Beam~1.\label{FitParaCorr}}
\end{figure}

Ten thousand beam imaging scans with 19 steps within the range of $\pm 4.5 \sigma_b$ are simulated with double-Gaussian beam shapes described in Eq.~\ref{doubleGauss}.
The beam-shape input parameters are varied within the following ranges
\begin{align}
\label{parameters}
w_{1,2} \in [0,1], \qquad \rho_{N,W} \in [-0.4, 0.4] , \\
\sigma_{N,x,y} \in [1.6, 2.0], \qquad  \sigma_{W,x,y} \in [2.0,2.6]\ ,
\end{align}  
where $w_{1,2}$ and $\rho_{N,W}$ are dimensionless and $\sigma_{N,x,y}$ and $\sigma_{W,x,y}$ are in arbitrary units.
Figure~\ref{FitParaCorr} shows a comparison of the true beam-shape parameters and the best-fit parameters extracted from the simultaneous fit of the four vertex distributions using the fit model derived in Equations~\ref{deriveFitModel3} to~\ref{deriveFitModel2}. 

\begin{figure}
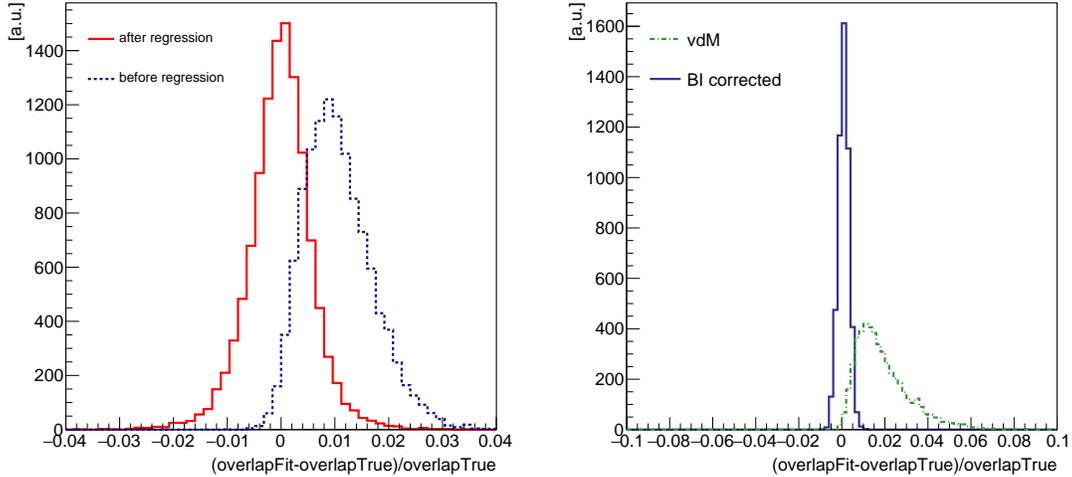

\centering
\includegraphics[width=208bp,height=201bp]{{{regressionPerformance}.pdf}}
\includegraphics[width=208bp,height=201bp]{{{vdmCorrection}.pdf}}
\caption{(Left) The relative difference of the true beam overlap and the reconstructed beam overlap using the beam imaging method before (blue dashed histogram) and after regression (red histogram) is shown. (Right) Relative difference of the true beam overlap and the reconstructed beam overlap using the standard vdM scan method (green dashed histogram) and the vdM scan method after corrections with the beam imaging method as shown in Eq.~\ref{BIhybrid}.\label{regression}}
\end{figure}

The beam overlap is then computed from the fit and compared to the true overlap calculated from the beam-shape input parameters, as shown in Figure~\ref{regression} (left plot, blue dashed histogram). A relative shift of approximately 1\% is observed. This difference can be viewed as a systematic bias of the method and is corrected for with the procedure developed in the next section.

Alternatively, the beam imaging method can be combined with the vdM scan method. For a given set of beam shapes a beam-imaging scan is performed to extract analytical models of the beam shapes. A vdM scan is then simulated to extract the bias introduced by non-factorization and used as correction factor for the vdM overlap integral estimates with the original beam shapes. 
\begin{align}
\label{BIhybrid}
O_I^{\mathrm{vdM}}\times\frac{O_I^{\mathrm{BI}}}{O_I^{\mathrm{BI,vdM}}} = O_I^{\mathrm{vdM, corrected}}
\end{align}  
The perfomance of this method is shown in Fig.~\ref{regression} right plot, where excellent closure for double-Gaussian beam shapes is observed.

\begin{figure}
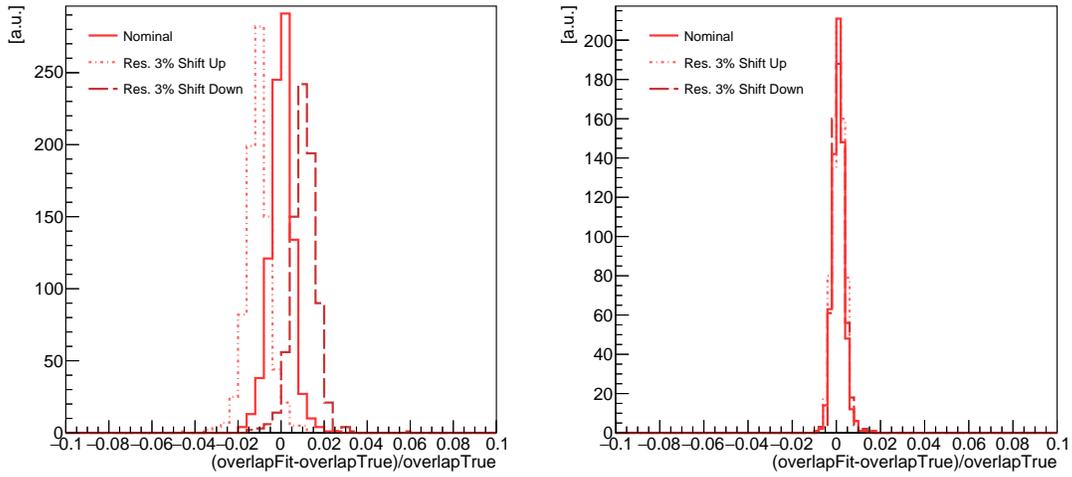

\centering
\includegraphics[width=208bp,height=201bp]{{{resolutionShift2}.pdf}}
\includegraphics[width=208bp,height=201bp]{{{VdMCorrectionResolution}.pdf}}
\caption{(Left) The relative difference of the true beam overlap and the reconstructed beam overlap using the beam imaging method with regression correction applying a 3\% shift up and down to the vertex position resolution. (Right) Relative difference of the true beam overlap and the reconstructed beam overlap using the standard vdM scan method after corrections with the beam imaging method as shown in Eq.~\ref{BIhybrid} applying a 3\% shift up and down to the vertex position resolution.\label{ResolutionShift}}
\end{figure}

The uncertainty on the vertex position resolution $V$ affects the estimate of the beam width parameters. As an example, a shift of 3\% is applied to the vertex resolution assumed in the simulation for which the impact is shown in Fig.~\ref{ResolutionShift} and translates into about $\pm$1\% uncertainty on the beam overlap integral after regression correction. The effective contribution to the total systematic uncertainty on the beam overlap estimate using beam imaging depends on the relative size of the vertex resolution compared to the preferably large beam size. Experimental effects in the vertex reconstruction and studies to measure the vertex position resolution are detailed in~\cite{Aaij:2014ida} for the beam-gas imaging method, where a systematic uncertainty of 1.2\% was determined.

As shown in Fig.~\ref{ResolutionShift} right, the beam overlap integral estimate using the vdM corrected method with Eq.~\ref{BIhybrid} is not observed to be affected by the uncertainties related to the vertex position.

\section{Bias Study and Regression}
\label{biasregression}
\begin{figure}[]
\centering
\includegraphics[width=215bp,height=211bp]{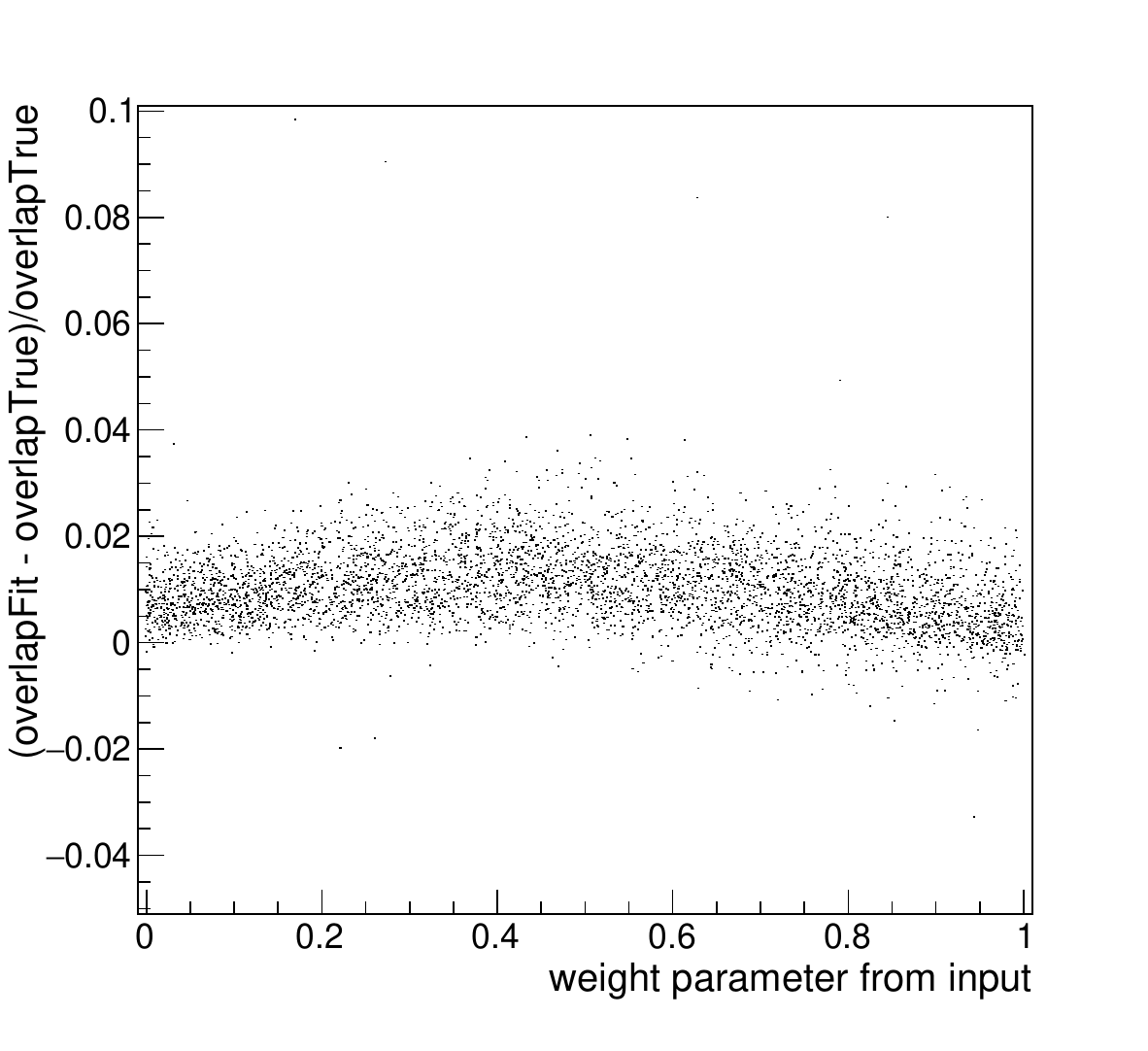}
\caption{Relative difference between the true overlap integral and the one obtained from beam imaging as a function of the weight parameter $w_1$ used as beam-shape input.\label{weightBias}}
\end{figure}

The on average 1\% shift in the reconstructed beam overlap estimated with the beam imaging technique and a dedicated correction method is discussed in this section.
The first source for the bias is due to the scan range used for simulation (about $\pm 4.5 \sigma_b$, i.e. $\pm 9$ in arbitrary units for the generated sample) which can in practice be limited by the available ring aperture in the experimental interaction region. While the fit model, i.e. Eq.~\ref{deriveFitModelX1}, assumes an integral over the full scan axis, the wide components of the beam shapes are only probed by part of the beam used for imaging. This results in an underestimation of the width parameters $\sigma_{i,W,x(y)}$ and thus an overestimate of the beam overlap integral. This bias disappears when the scan range is increased. In addition, a correlation is observed between the bias on the overlap integral and the input parameters $w_i$, with the bias reaching a maximum at values $w_i\approx 0.5$, as shown in Fig.~\ref{weightBias}. No other significant correlations between other parameters and $w_{1,2}$ are observed and this bias is therefore interpreted as a feature of the beam imaging method.

The method proposed to correct for the bias utilizes the regression functionalities based on neural networks of the TMVA package~\cite{Hocker:2007ht}. A regression for the beam overlap is trained using the fitted beam shape parameters as input variables targeting the true beam overlap integral computed from the input beam shape parameters
\begin{equation}
f^{\mathrm{regression}}: S(P^{\mathrm{fit}}) \longrightarrow S(O_I^{\mathrm{true}}),
\end{equation}
where $S(P^{\mathrm{fit}})$ is the set of fitted parameters listed in Eq.~\ref{parameters} and $S(O_I^{\mathrm{true}})$ is the set of beam overlap integrals calculated from the input parameters of the bunch proton densities.
The sample of ten thousand simulated beam imaging scans is split into two random subsets of samples with five thousand scans each. One sample is used to train the regression and the other one is used to evaluate the performace of the regression to prevent training bias. The result is shown in Figure~\ref{regression}, red histogram. Compared to the uncorrected beam overlap, the blue dashed histogram, the regression method improves the resolution to 0.6\% and successfully removes the bias.
For this correction method to be applicable experimentally, the validity of the fit model needs to be evaluated first to generate a suitable training sample from simulation. If more than one model fits the data, it is suggested to generate training samples for each model and estimate an uncertainty from the comparison accordingly.
For the application of the non-factorization correction shown in Eq.~\ref{BIhybrid} and Fig.~\ref{regression}, no bias is observed.

\section{Conclusions}

We discussed a method to reconstruct two-dimensional proton-bunch densities using vertex distributions recorded in LHC beam-beam scans. 
The $x$-$y$ correlations in the beam shapes are studied and an alternative luminosity calibration technique is introduced. 
As shown in Section~\ref{vdmCorrStudy}, correlations in $x$-$y$ can lead to significant biases on the luminometer visible cross section estimate. 
The beam imaging method presented in this paper allows to directly measure the underlying $x$-$y$ correlations, and thereby determine the two-dimensional beam-overlap integral. The method is directly applicable for transverse beam shapes that can be analytically convolved with vertex position resolution models. For arbitrary beam shapes, where numerical convolution methods have to be utilized, limitations due to computational performance arise. In Section~\ref{2dbeamimaging}, we propose a fit model for two-dimensional vertex distributions in the transverse plane of a detector accumulated during beam imaging scans, where one beam is kept fixed and the other one is moved in $x$ and $y$. We evaluate the method on a sample of simulated beam imaging scans with double-Gaussian beam shapes. After applying a dedicated correction based on a neural network regression technique the beam overlap integral is reconstructed to good precision over a large range of parameters describing non-factorization. Deployed under experimental conditions, the impact on the method introduced by effects such as beam-orbit drift, spatial vertex reconstruction resolution, and beam-beam effects have to be investigated. Alternatively, the beam imaging method can be combined with the standard vdM scan method to account for non-factorization effects, where excellent performance is shown.

\section{Acknowledgments}
We are grateful for the support of Catherine Medlock by the MIT-MISTI program and of Jakob Salfeld-Nebgen by the German Research Foundation. We would like to thank Colin Barschel and Marco Zanetti for fruitful discussions in the initial phase of this project.


\end{document}